\documentclass[prb,showpacs,floatfix,preprint]{revtex4}

\usepackage{epsfig,graphicx} 
\usepackage{dcolumn} 
\usepackage{bm}
\usepackage{float}

\def\fvo{FeVO$_4$}
\def\nvo{Ni$_3$V$_2$O$_8$}
\def\rhov{{\mbox{\boldmath{$\rho$}}}}
\def\tauv{{\mbox{\boldmath{$\tau$}}}}
\def\Lambdav{{\mbox{\boldmath{$\Lambda$}}}}
\def\Thetav{{\mbox{\boldmath{$\Theta$}}}}
\def\Psiv{{\mbox{\boldmath{$\Psi$}}}}
\def\Phiv{{\mbox{\boldmath{$\Phi$}}}}

\begin{document} 

\title{Magnetic structure and magnetoelectric coupling in bulk and
thin film FeVO$_4$} 

\author{A. Dixit$^1$, G. Lawes$^1$, A.B. Harris$^2$} 

\affiliation{$^1$Department of Physics and Astronomy, Wayne State
University, Detroit, MI 48201\\
$^2$Department of Physics and Astronomy, 
University of Pennsylvania, Philadelphia, PA 19104} 

\date{\today} 

\begin{abstract} 
We have investigated the magnetoelectric and magnetodielectric response in
\fvo, which exhibits a change in magnetic structure coincident with ferroelectric
ordering at $T_{N2}$$\approx$15 K.  Using symmetry considerations, we
construct a model for the possible magnetoelectric coupling in this system,
and present a discussion of the allowed spin structures in \fvo.  Based on this model,
in which the spontaneous polarization is caused by a
trilinear spin-phonon interaction, we experimentally explore the magnetoelectric coupling in \fvo\,
thin films through measurements of the electric field induced shift of the
multiferroic phase transition temperature, which exhibits an increase of
0.25 K in an applied field of 4 MV/m.  The strong spin-charge coupling in
\fvo\, is also reflected in the significant magnetodielectric shift, which
is present in the paramagnetic phase due to a quartic spin-phonon interaction
and shows a marked enhancement with the onset of magnetic order
which we attribute to the trilinear spin-phonon interaction.
We observe a clear magnetic
field induced dielectric anomaly at lower temperatures, distinct from the
sharp peak associated with the multiferroic transition, which we tentatively
assign to a spin reorientation cross-over.  We also present a magnetoelectric
phase diagram for \fvo.
\end{abstract} 


\maketitle

\section{INTRODUCTION}

Magnetoelectric multiferroics, insulating magnets exhibiting simultaneous
magnetic and ferroelectric order, are widely investigated in large part
due to their potential applications for developing novel devices, including
magnetic sensors and multistate memory, among others [\onlinecite{ABHGL,saf2006}].
The cross-control
of these distinct order parameters, such as adjusting the magnetization
using an applied electric field or vice-versa, is expected to provide an
extra degree of freedom in developing new types of spin-charge coupled
devices, such as voltage switchable magnetic memories [\onlinecite{Eeren2007}].
Additionally,
there are a number of fundamental materials questions surrounding the 
development of multiferroic order. Magnetic and ferroelectric order are
generally contraindicated in the same phase, as ferromagnetism in 
transition metal systems typically requires partially filled d-orbitals
while ferroelectric distortions are promoted in a d$^0$ electronic 
configuration[\onlinecite{Hill2000}]. Despite this apparent restriction,
a rather large number of
single phase systems have been identified as magnetoelectric 
multiferroics [\onlinecite{TK2005,Hur2004,Lawes2005}]. A number of 
microscopic mechanisms
have been proposed for the development of multiferroic order, including,
a magnetic Jahn-Teller distortion[\onlinecite{JT}] for 
TbMn$_2$O$_5$,[\onlinecite{LCC2004}] bond and site ordering having
distinct centers of inversion symmetry,[\onlinecite{Efremov2004}]
a microscopic mechanism leading to a spin-current 
interaction[\onlinecite{Katsura2005}],
the Dzyloshinskii-Moriya interaction,[\onlinecite{SergienkoPRB}]
a general anisotropic exchange striction,[\onlinecite{HYAE}]
a spin-phonon interaction,[\onlinecite{Tackett}] and
a strain induced ferroelectricity.[\onlinecite{RABE3}]

Phenomenologically,  magnetically-induced ferroelectric order developing
in systems having multiple magnetic phases can be understood by considering
a trilinear term in the magnetoelectric free energy, F$_{ME}$, coupling
the electric polarization with two distinct order parameters 
 $\sigma_1$ and $\sigma_2$ which together break inversion symmetry so that 
F$_{ME}\propto$P$\sigma_1\sigma_2$[\onlinecite{ABHGL,Lawes2005,MKPRL,
ABHJAP,ABHPRB}].  Since the free energy must
transform as a scalar, there are strong symmetry restrictions on the
allowed representations for $\sigma_1$ and $\sigma_2$; in particular,
the product $\sigma_1({\bf q})\sigma_2({\bf q})^*$ must be antisymmetric 
under spatial inversion.  This trilinear coupling also predicts
electric field control of a magnetic order 
parameter[\onlinecite{ABHGL,Kharel2009}].  A general discussion of
the symmetry of the magnetoelectric coupling in multiferroics is considered
for the specific case of \fvo\, in the following section.  Investigations on 
multiferroic \nvo\, thin films, in which such a trilinear coupling is 
believed to be responsible for the multiferroic order[\onlinecite{ABHGL,Lawes2005}], have established
that the multiferroic transition temperature can be varied through the
application of either (or both) magnetic and electric
fields [\onlinecite{Kharel2009}], confirming the strong coupling
between magnetic and dielectric degrees of freedom.   Higher order magnetoelectric 
coupling terms quadratic in both magnetic and ferroelectric terms will
give rise to magnetization induced shift in the dielectric
response[\onlinecite{RABE1}].
Such coupling has been investigated both theoretically and experimentally
in a range of materials including Mn$_3$O$_4$[\onlinecite{Tackett}],
CoCr$_2$O$_4$[\onlinecite{LawesMelot}], 
BaMnF$_4$[\onlinecite{Samara1976,Fox1977,Fox1979,Scott1977}], and
SeCuO$_3$ and TeCuO$_3$[\onlinecite{Lawes2003}].  Because this 
magnetodielectric coupling is also expected to depend strongly on the
symmetry of the magnetic phase, it has been suggested that changes in 
this coupling may be used to probe changes in the ordered 
spin structure[\onlinecite{Tackett}].

Triclinic iron vanadate, \fvo, has recently been identified as a multiferroic
system having the P$\overline{1}$ space group [\onlinecite{Dixit2009,Daoud2009,Kundys2009}].
Magnetic, thermodynamic, and neutron diffraction studies on \fvo\, 
single crystal and ceramic samples have shown that \fvo\, transitions from a paramagnetic
phase into a collinear incommensurate (CI) phase at 
$T_{N1}$=22 K and then into non-collinear incommensurate (NCI) phase at
$T_{N2}$=15 K.[\onlinecite{Dixit2009,Daoud2009,Kundys2009,He2008}]
Ferroelectric order in \fvo\,develops in this non-collinear 
spiral magnetic phase.  The onset of ferroelectric order with
the development of a second magnetic phase suggests that a symmetry-based
approach may be useful in exploring the multiferroic properties in this system.
We present a full Landau theory for this system, specifically considering the 
allowed magnetoelectric coupling terms.
Our result is that a nonzero induced spontaneous polarization
${\vec P}$ requires having a magnetic spiral[\onlinecite{MOST}]
described by two order parameters which are out of phase with respect to 
one another.[\onlinecite{Lawes2005,MKPRL,ABHGL,ABHPRB}]  
In this low symmetry structure
there are no restriction on the orientation of ${\vec P}$ based on symmetry
arguments, unlike the majority of similar magnetically-induced
multiferroics[\onlinecite{ABHGL,Lawes2005,MKPRL,MOST}].  

This paper is organized as follows.  In Sec. II we present a symmetry
analysis of FeVO$_4$ based on Landau theory.  Here we analyze the
symmetry of the magnetoelectric interaction.  In Sec. III we
present the results of a number of experiments designed to probe
the structure of these magnetoelectric interactions.  In Sec. IV
we briefly summarize our results.

\section{Landau Theory}

Motivated by this general discussion of the possibility of magnetically
driven ferroelectric order in \fvo, we now present a Landau theory for
\fvo\, with some details of the construction relegated to the
the Appendix.  As discussed in detail in Ref. [\onlinecite{ABHPRB}],
the Fourier transform of the spin ordering is proportional to the
critical eigenvector of the
inverse susceptibility matrix at the ordering wave vector $\vec q$.
In the Appendix we analyze the constraint of 
 spatial inversion in 
the P$\overline{1}$ space group of the paramagnetic phase,
with the following results.  The \fvo\, structure consists of six
$S$=5/2 Fe$^{3+}$ spins in the unit cell at locations $\tauv_n$. For
$n=1,2,3$, $-\tauv_n=\overline \tauv_n=\tau_{n+3}$ and
$-\tauv_{n+3}=\overline \tauv_{n+3}=\tauv_n$. Then
inversion symmetry (${\cal I}$) implies that spin Fourier transform obeys
\begin{eqnarray}
{\cal I} \vec S(\vec q, \tau) = \vec S (\vec q , \overline \tau)^* \,
\end{eqnarray}
where, as defined in the Appendix, $\vec S (\vec q, \tau)$ is
the spatial Fourier transform of the thermally averaged spin operator.
As explained in the Appendix, this
relation implies that the spin distribution is inversion-symmetric
about some origin. We find that
inversion symmetry implies that
\begin{eqnarray}
[S_x(1), S_y(1),S_z(1), S_x(2),S_y(2) \dots S_z(6)] &=&
\sigma_n [ x_1^*, y_1^*, z_1^*, x_2^*,y_2^*, ... z_6^*] \ 
\end{eqnarray}
where all the components are complex valued with
$x_{\overline n}=x_n^*$, $y_{\overline n}=y_n^*$, $z_{\overline n}=z_n^*$,
are normalized by $\sum_{n=1}^6 [|x_n|^2 + |y_n|^2 + |z_n|^2] =1$,
and the wave vector argument is implicit.  The amplitude
$\sigma_n (\vec q)$ is the complex valued magnetic order parameter, which obeys
\begin{eqnarray}
{\cal I} \sigma_n(\vec q)=\sigma_n(\vec q)^*=\sigma_n(-\vec q)\ .
\label{ISYM} \end{eqnarray}
As noted in the Appendix, this relation implies that each $\sigma_n$
is inversion invariant about a lattice point (which depends on $n$)
where the order parameter wave has its origin.
As the temperature is lowered one passes from the paramagnetic phase 
into a phase with an order parameter $\sigma_1(\vec q)$ and
then, at a lower temperature into a phase where two order
parameters $\sigma_1(\vec q)$ and $\sigma_2(\vec q)$ are nonzero,
both of which obey Eq. (\ref{ISYM}), but which have 
different centers of inversion symmetry.

The total magnetoelectric free energy, $F_{\rm ME}$ can be written as
\begin{eqnarray}
F_{\rm ME} &=& F_{\rm M} + F_{\rm E} + V \ ,
\end{eqnarray}
where $F_{\rm M}$ is the purely magnetic free energy,
$F_{\rm E}$ is the dielectric potential which we approximate as
$F_{\rm E} = (1/2) \chi_E^{-1} {\bf P}^2$, where $\chi_E$ is
the dielectric susceptibility (whose crystalline anisotropy is neglected),
and to leading order in $\sigma_n$, the magnetoelectric coupling term is given by:
\begin{eqnarray}
V &=& \sum_{n,m=1}^2 \sum_\gamma 
[a_{n,m,\gamma} \sigma_n(\vec q) \sigma_m(\vec q)^*
+ a_{n,m,\gamma}^* \sigma_n(\vec q)^* \sigma_m(\vec q)] P_\gamma \  ,
\end{eqnarray}
where $n$ and $m$ label order parameter modes and $\gamma$ labels the
Cartesian component of $\vec P$.  Terms linear in $\sigma_n$ are prohibited
because they are not time reversal invariant and also can not conserve wave vector.
The magnetoelectric interaction $V$
has to be inversion invariant and the appendix shows that
the $a$ coefficients are pure imaginary, so that
\begin{eqnarray}
V &=& i \sum_\gamma r_\gamma [\sigma_1(\vec q) \sigma_2(\vec q)^* -
\sigma_1(\vec q)^* \sigma_2(\vec q)] P_\gamma =
2 \sum_\gamma r_\gamma |\sigma_1(\vec q) \sigma_2(\vec q)|
\sin(\phi_2 - \phi_1) P_\gamma \ ,
\end{eqnarray}
where $\sigma_n(\vec q) = |\sigma_n(\vec q)| \exp (i \phi_n)$.
There is no restriction on the direction of the spontaneous
polarization, so that all components of $\vec P$ will be nonzero.
However, if the magnetic structure is a spiral, then the arguments of
Mostovoy[\onlinecite{MOST}] might be used to predict the approximate
direction of $\vec P$.  The result of Eq. (6)
 is quite analogous
to that for Ni$_3$V$_2$O$_8$[\onlinecite{Lawes2005}]
or for TbMnO$_3$[\onlinecite{MKPRL}], in that it requires the presence
of two modes $\sigma_1 (\vec q)\equiv \exp(i \phi_1) |\sigma_1(\vec q)|$ and
$\sigma_2 (\vec q)\equiv \exp(i \phi_2) |\sigma_2(\vec q)|$ which are out
of phase with one another: $\phi_1 \not= \phi_2$.  
Then the order parameter wavefunctions have different
origins and will therefore break inversion symmetry.

If, as stated in Ref. \onlinecite{Daoud2009}, the eigenvector is {\it not}
inversion invariant as implied by Eq. (\ref{ISYM}), then
one would conclude that the magnetic ordering transition is not continuous.
However, the most likely scenario is that the ordering transitions are
continuous and that the spin distribution for each $\sigma_n(\vec q)$ is inversion symmetric as
obtained in this derivation.  The acentric distribution found in 
Ref. \onlinecite{Daoud2009} differs only slightly from being inversion symmetric
for reasons that are obscure.[\onlinecite{PGRPC}]  

\section{Magnetoelectric Interactions (Experimental)}

\subsection{Sample Synthesis and Structural Characterization}

Motivated by Eq. (6), which predicts that the magnetic structure
defined by $\sigma_1(\vec q)$ and $\sigma_2 (\vec q)$ is coupled to the
electric polarization $P$, we experimentally investigated the nature of the higher order
magnetoelectric coupling in \fvo. Bulk single phase polycrystalline iron
vanadate (\fvo) ceramic samples were prepared using standard solid state
reactions.  Because  Eq. 6 predicts that all components of the
polarization vector are nonzero, and previous measurements on
ceramic \fvo\, have found clear evidence for multiferroic
behaviour [\onlinecite{Dixit2009}] we focused our study on polycrytalline
samples. A stoichiometric ratio of iron oxide (Fe$_2$O$_3$) and vanadium
pentaoxide (V$_2$O$_5$) solid solutions were thoroughly mixed and ground
to produce a homogeneous mixture. This mixture was slowly heated to 600$^{\rm o}$C
 for 4 hours in air. Intermediate grindings followed by thermal
annealing in air were repeated several times to complete the solid state
reaction and ensure a fully reacted and uniform composition. This homogeneous
solid solution was finally annealed in air at 800$^{\rm o}$C for 4 hours,
yielding a yellowish brown powder identified as a single phase iron vanadate
by X-ray diffraction and Raman spectroscopy.

In order to apply large electric fields to \fvo\, we also prepared thin
film samples. These were fabricated from a phase pure stoichiometric iron
vanadate target. The \fvo\,powder used for the sputtering target was 
prepared by the method described above. Approximately 30 g of \fvo\,powder
was mixed with 15 mL of 2 mole percent polyvinyl alcohol as a binder. The
dried powder was pressed into a circular disc having a diameter of 
approximately 50 mm with a thickness of roughly 3.5 mm followed by air 
annealing at 600$^{\rm o}$C for 4 hours to burn off the residual organics.
A final thermal annealing was done at 800$^{\rm o}$C for 4 hours to 
produce the dense pellet used for the sputtering target. \fvo\,films were 
deposited at room temperature using RF magnetron sputtering onto conducting
silicon substrates.  The working pressure was held at 1.5x10$^{-2}$ torr,
with the atmosphere consisting of a mixture of approximately 1.5x10$^{-3}$ torr
partial pressure of oxygen as the reactive gas and approximately 1.35x10$^{-2}$
torr partial pressure of argon as the sputtering gas. These as 
deposited films, prepared over a time of 4 hours, were amorphous.  After
air annealing at 700$^{\rm o}$C for 4 hours the films were indexed as
single phase polycrystalline \fvo. 

We investigated the structural, magnetic, and electronic properties of
these samples using a number of different techniques. We used a Rigaku
RU200 powder X-ray diffractometer and Horiba Triax Raman spectrometer to
study the crystalline structure of these samples. We used a Hitachi scanning
electron microscope (SEM) to investigate the surface morphology of the thin
film samples and an associated energy dispersive X-ray (EDX) assembly to
probe the chemical composition of both samples. We measured the temperature
dependent magnetization of the powder sample using a Quantum Design MPMS
SQUID magnetometer, although the very small magnetic anomalies associated
with the transitions could not be clearly distinguished from background in
the thin films samples. We conducted temperature and field dependent 
dielectric and pyrocurrent measurements using the temperature and field
control provided by a Quantum Design PPMS system used in conjunction with
an Agilent 4284A LCR meter and a Keithley 6517 electrometer. These 
measurements were done on a cold pressed pellet of bulk \fvo\,with the top
and bottom electrodes fashioned using silver epoxy and on the \fvo\,thin
films with room sputtered gold (Au) used as the top electrode and the Si
substrate serving as the bottom electrode. 

\begin{figure}
\centering\includegraphics[width=10.5cm]{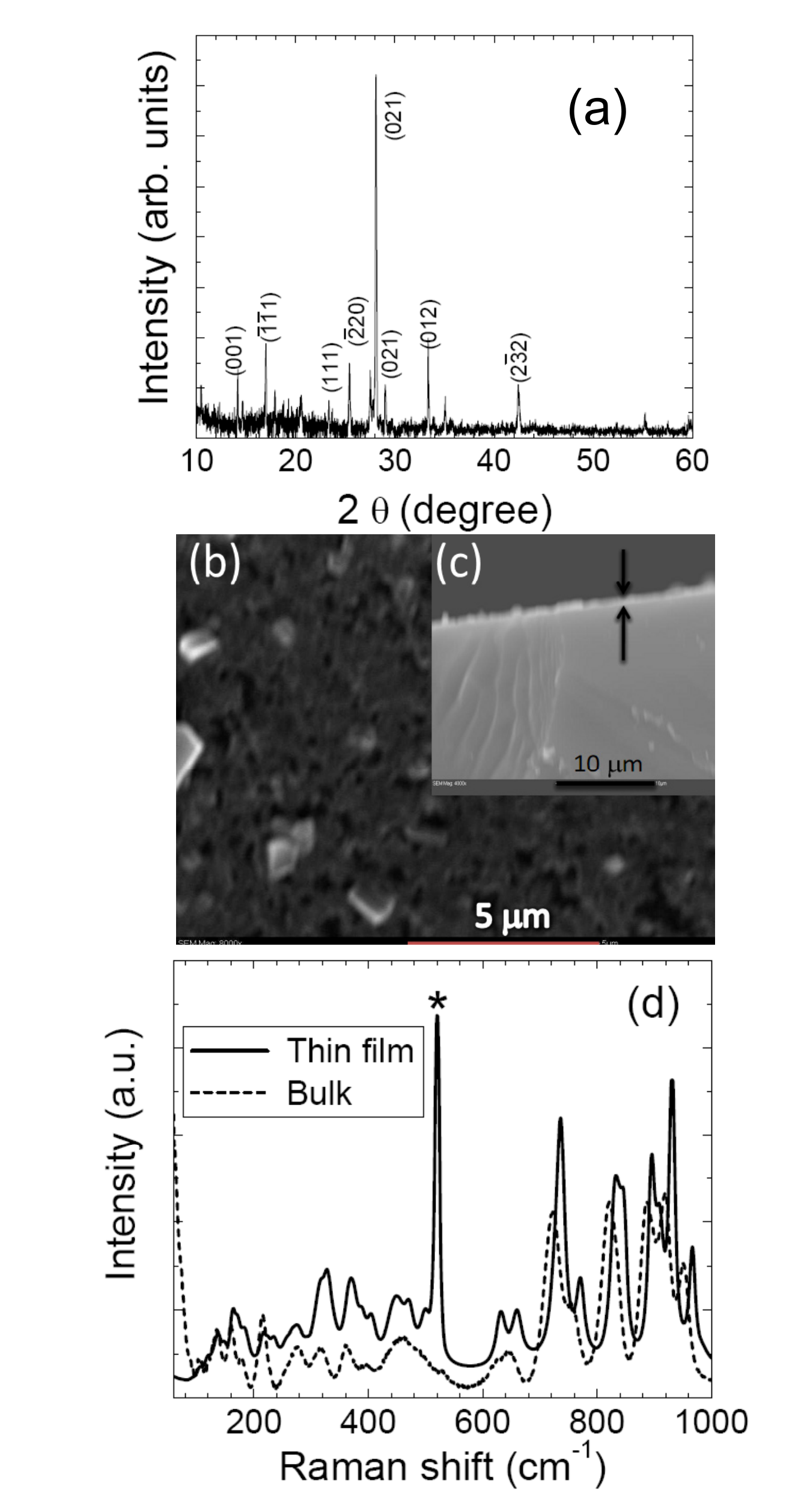}\\
\caption{(a) $\theta$-2$\theta$ x-ray diffraction (XRD) pattern of \fvo\, thin film, 
(b) Surface scanning electron micrograph of \fvo\, thin film, 
(c) Cross-sectional SEM image of \fvo, thin film and 
(d) room temperature Raman spectrum on \fvo\, bulk powder and thin film samples.
The peak at 560 cm$^{-1}$ (indicated by an asterisk) arises from
the silicon substrate}
\label{fig:Fig1}
\end{figure}
 
The structure of the ceramic \fvo\,sample was practically identical to that
previously presented for a bulk sample prepared using a different 
technique [\onlinecite{Dixit2009}]. The X-ray diffraction (XRD) pattern for the \fvo\,thin
film is shown in Fig 1(a). These diffraction peaks are consistent with
the expected XRD pattern for \fvo\,[JCPDS no.  38-1372]. The surface 
morphology the thin film sample is shown in Fig 1(b).  This SEM micrograph
indicates that the film consists of grains with various orientations, as
well as a number of pinhole defects. We calculated the thickness of these
thin films to be roughly 200 nm, using the cross-sectional SEM micrograph,
Fig 1(c).  This value is very consistent with estimates from well defined
interference fringes observed in reflection spectra (not shown). EDX 
analysis of both the bulk and thin film samples show a 1:1 iron to vanadium
ratio. We carried out room temperature Raman vibrational spectroscopy to
further probe the microstructures of both bulk and thin films. The 
identification of Raman active modes and their detailed temperature 
dependent analysis on bulk \fvo\,sample is discussed elsewhere [\onlinecite{Dixit2009}].
Here we plot the room temperature Raman spectrum of both bulk and thin
film \fvo\,in Fig1(d). We are able to identify all the Raman active modes
for thin films, which are observed in bulk \fvo\,[\onlinecite{Dixit2009}] with a small shift
in the Raman peaks for the thin films.  The Raman peak arising from the
silicon substrate is indicated by an asterisk.

\subsection{Temperature Dependent Dielectric Measurements on \fvo\, Ceramic}

\begin{figure}
\centering\includegraphics[width=9.5cm]{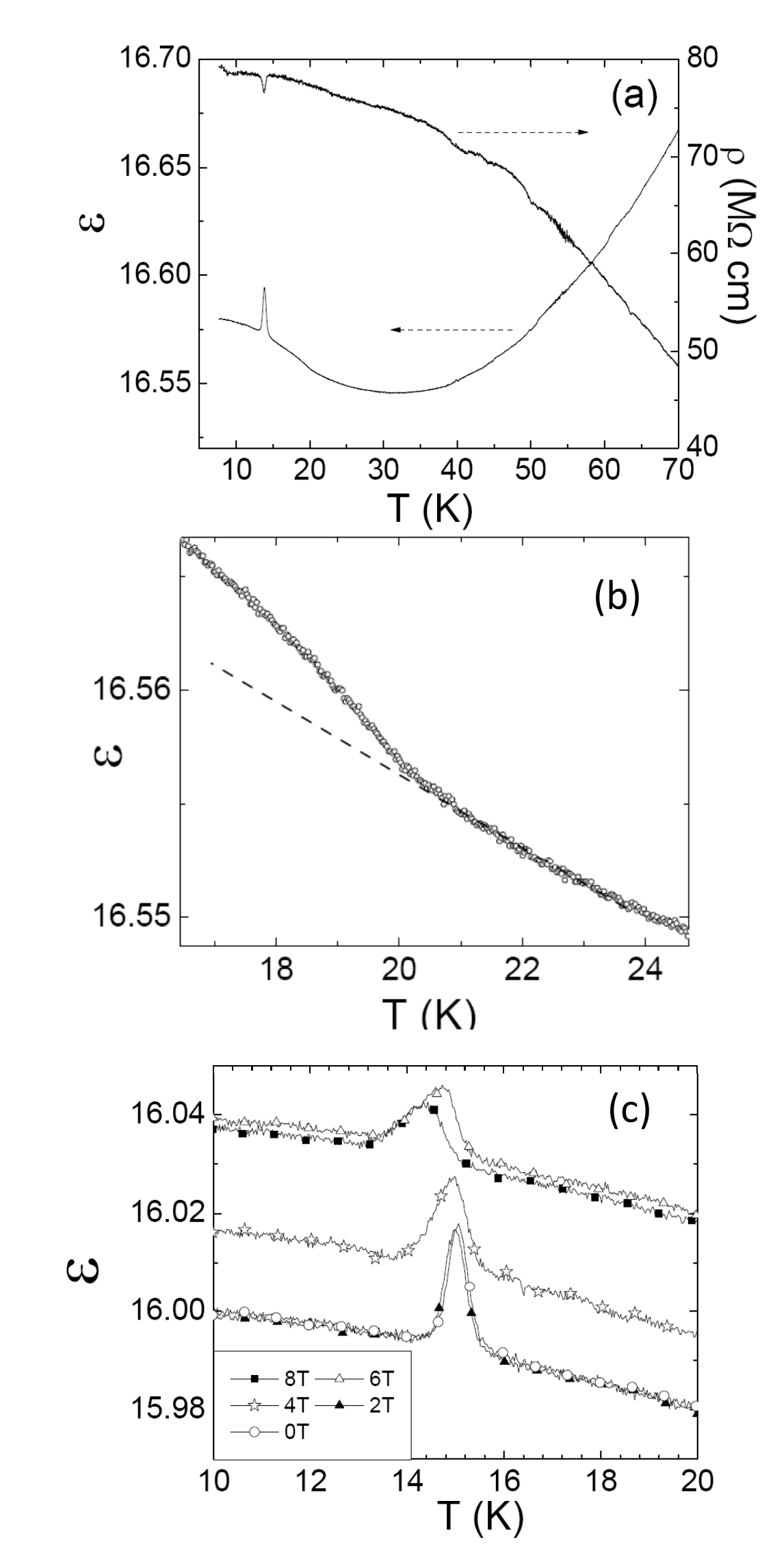}\\
\caption{(a) Zero field temperature dependent dielectric constant (left axis) 
and resistivity (right axis) for ceramic \fvo\, sample, 
(b) Temperature dependence of the dielectric constant near the 
magnetic ordering temperature, and 
(c) Temperature dependence of dielectric constant at $H$= 0, 20, 40, 60, and 80 kOe.  
The dashed line in (b) is a guide to the eye.}
\label{fig:Fig2}
\end{figure}

The temperature dependent magnetization for the bulk \fvo\,sample (not
shown) was practically identical to that measured previously on a different
ceramic sample prepared using a different technique [\onlinecite{Dixit2009}].  In particular,
the magnetization showed the usual two anomalies associated with the two
incommensurate transitions in this system.  We plot the zero-field 
dielectric constant and resistivity for bulk \fvo\,over a broad range of
temperatures in Fig. 2(a).  The dielectric constant exhibits a sharp peak
near $T_{N2}$, arising from the development of ferroelectric order in the
incommensurate spiral magnetic phase.   As shown in Fig. 2(a), above 35 K,
the dielectric constant for \fvo\,shows a gradual decrease on cooling,
typical of many insulating materials [\onlinecite{Ramirez2000}].
  Below roughly 30 K, the dielectric
constant increases smoothly with further cooling.  Since the resistivity
of \fvo\,increases monotonically with decreasing temperature (except for
a small anomaly at $T_{N2}$), also shown in Fig. 2(a), we attribute this
increase in the dielectric constant to a quartic magnetoelectric
coupling, $V_4$.  
Although \fvo\,does not order magnetically
until cooled below $T_{N1}$=22 K, heat capacity measurements suggest the
presence of short range spin correlations developing well above this 
temperature [\onlinecite{Dixit2009}].  It has been suggested in a number of other systems,
including TeCuO$_3$ [\onlinecite{Lawes2003}] and 
Mn$_3$O$_4$ [\onlinecite{Tackett}], that short-range magnetic
correlations can produce magnetodielectric corrections; we propose that the
same mechanism is responsible for the non-monotonic temperature dependence
of the dielectric constant of \fvo\,in the paramagnetic phase.

The fourth order magnetoelectric coupling contains terms quadratic in
$\vec P$ and $\sigma_n(\vec q)$.  If $\sigma_1(\vec q)$ is the 
order parameter that develops at $T_{N1}$, then this coupling is 
probably dominated by
 $\lambda|\vec P|^2|\sigma_1(\vec q)|^2$ at high temperatures.
The finite spin correlations developing 
above $T_{N1}$ cause $\langle |\sigma_1(\vec q)|^2\rangle$
to be nonzero so that the coupling term is in effect
$a |\vec P|^2$,
where $a= \lambda \langle|\sigma_1(\vec q)|^2\rangle$.
This term
produces a shift in the dielectric constant in the 
paramagnetic phase, as seen in Fig. 2(a) below approximately 30 K.  Below
$T_{N1}$, approximately 22 K, when $\sigma_1(\vec q)$ 
acquires a finite expectation value,
a trilinear coupling term $a\vec P<\sigma_1(\vec q)>\sigma_2(\vec q)^*$ is
allowed.  This term will lead to mode mixing so that the critical mode
approaching $T_{N2}$ is not $\sigma_2(\vec q)$ but 
$(\sigma_2(\vec q)+\rho\vec P)$
where $\rho$ is of order $a\langle \sigma_1(\vec q) \rangle$ 
[\onlinecite{ABHPRB}]. 
Then the
divergence in this variable as $T_{N2}$ is approached will
lead to a simultaneous divergence (with a very much
reduced amplitude) in the observed dielectric constant.
This mode mixing is therefore expected to lead
to a slight increase in the magnetodielectric shift below $T_{N1}$ [\onlinecite{ABHPRB}].
This is seen clearly in Fig. 2b, where the dashed line shows the extrapolation
of the magnetodielectric shift from $T>T_{N1}$ to $T<T_{N1}$, an
extrapolation that does not include any 
corrections arising from the trilinear magnetoelectric term.

\subsection{Magnetic Field Dependent Dielectric Measurements on \fvo\, Ceramic}

To further investigate spin-charge coupling in \fvo, we plot the temperature
dependent dielectric constant measured at different magnetic fields in
Fig 2(c). We find that the dielectric anomaly signaling the onset of 
ferroelectric order shifts to lower temperatures with increasing magnetic
field, with the reduction in transition temperature reaching  0.7 K in a
magnetic field of $H$=80 kOe.  This result is expected, as the ferroelectric
order producing the dielectric anomaly is associated with the incommensurate spiral
transition, which typically show a reduction in transition temperature in applied
magnetic fields. 

\begin{figure}
\centering\includegraphics[width=12cm]{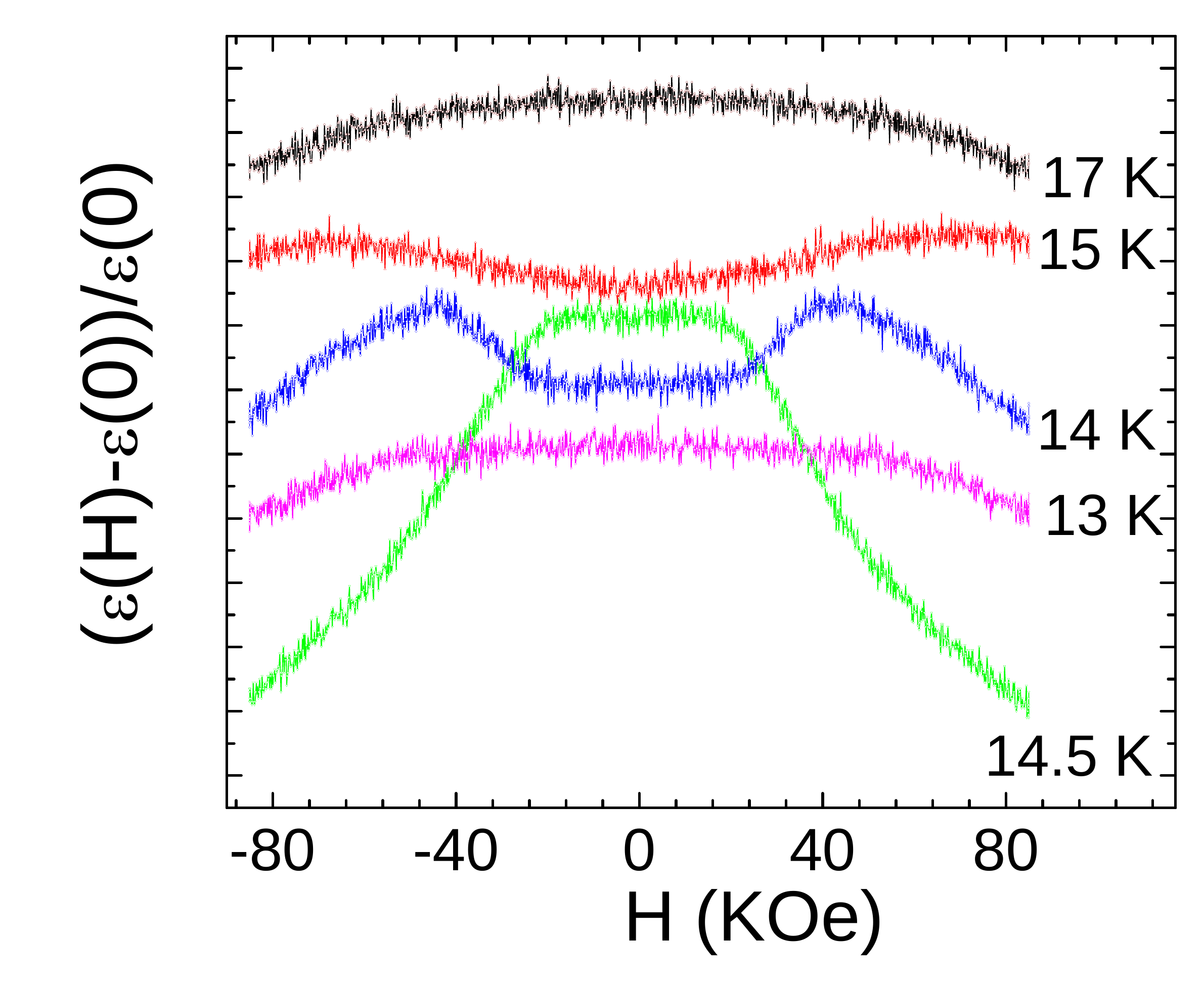}\\
\caption{(color on-line)Magnetic field dependence of the relative change in the 
dielectric constant for ceramic \fvo\, at different temperatures with vertical offset included for clarity.}
\label{fig:Fig3}
\end{figure}

We conducted additional measurements of the dielectric response of the bulk
sample while sweeping the magnetic field at fixed temperature.  These 
results are shown in Fig. 3, plotted as $\Delta\epsilon(H)/ \epsilon(H=0)$
versus $H$ with data measured at different temperatures offset vertically for
clarity.  At $T$=17 K, which is intermediate between $T_{N1}$ and $T_{N2}$,
there is a small negative magnetocapacitance, with the dielectric constant
being reduced by approximately 0.03\% in a field of $H$=80 kOe.  As the 
temperature approaches the multiferroic transition at $T_{N2}$, the 
magnetodielectric coupling shows qualitative changes.  By $T$=15 K the 
magnetocapacitive shift is positive for small fields, with a shift in 
dielectric constant on the order of 0.02\% at high magnetic fields.  
The magnetocapacitive response is maximal near $T$=14.5 K, with the 
dielectric constant being reduced by approximately 0.1\% in a field of
$H$=80 kOe.  At still lower temperatures the magnitude of the magnetocapacitive
shift becomes smaller.

Perhaps the most dramatic feature in the isothermal magnetocapacitance curves
presented in Fig. 3 is the presence of clear maxima, which vary as a function
of temperature and magnetic field.  These maxima appear first at small fields
at $T$=14.5 K, then shift to larger fields as the temperature is reduced.
We believe that these anomalies do not reflect the suppression of the 
multiferroic transition temperature in a magnetic field, as discussed in the
context of Fig 2(b).  These isothermal dielectric anomalies persist to 
temperatures 2 K or 3 K below $T_{N2}$, while the maximum suppression of
$T_{N2}$ was only 0.7 K over the field range studied, as determined from
the measurements in Fig. 3.  We propose that this dielectric anomaly may
indicate a spin-reorientation transition in \fvo.  The magnetodielectric coupling
is expected to depend on the symmetry of the magnetically ordered state
[\onlinecite{Lawes2003,ABHPRB}],
so a field induced spin reorientation crossover could potentially produce
the low temperature dielectric anomalies observed in Fig. 3.  Similar magnetic
field-induced dielectric anomalies have been observed in other materials
including Mn$_3$O$_4$ [\onlinecite{Tackett}], although the 
specific mechanisms responsible
remain unclear.  One possibility is that the external magnetic field serves
to reduce the slight geometrical frustration present 
in \fvo\,[\onlinecite{Daoud2009}], allowing
a different spin structure to emerge.
Alternatively, the spin orientation could be a spin-flop
transition as seen in TbMnO$_3$.[\onlinecite{Aliouane2006}]  
We note, however, that \fvo\,remains
ferroelectric at high magnetic fields [\onlinecite{Dixit2009}], 
so the modified spin structures
would still need to transform as defined by Eq. 3.

\subsection{Magnetoelectric Coupling in \fvo\, Thin Films}

\begin{figure}
\centering\includegraphics[width=12cm]{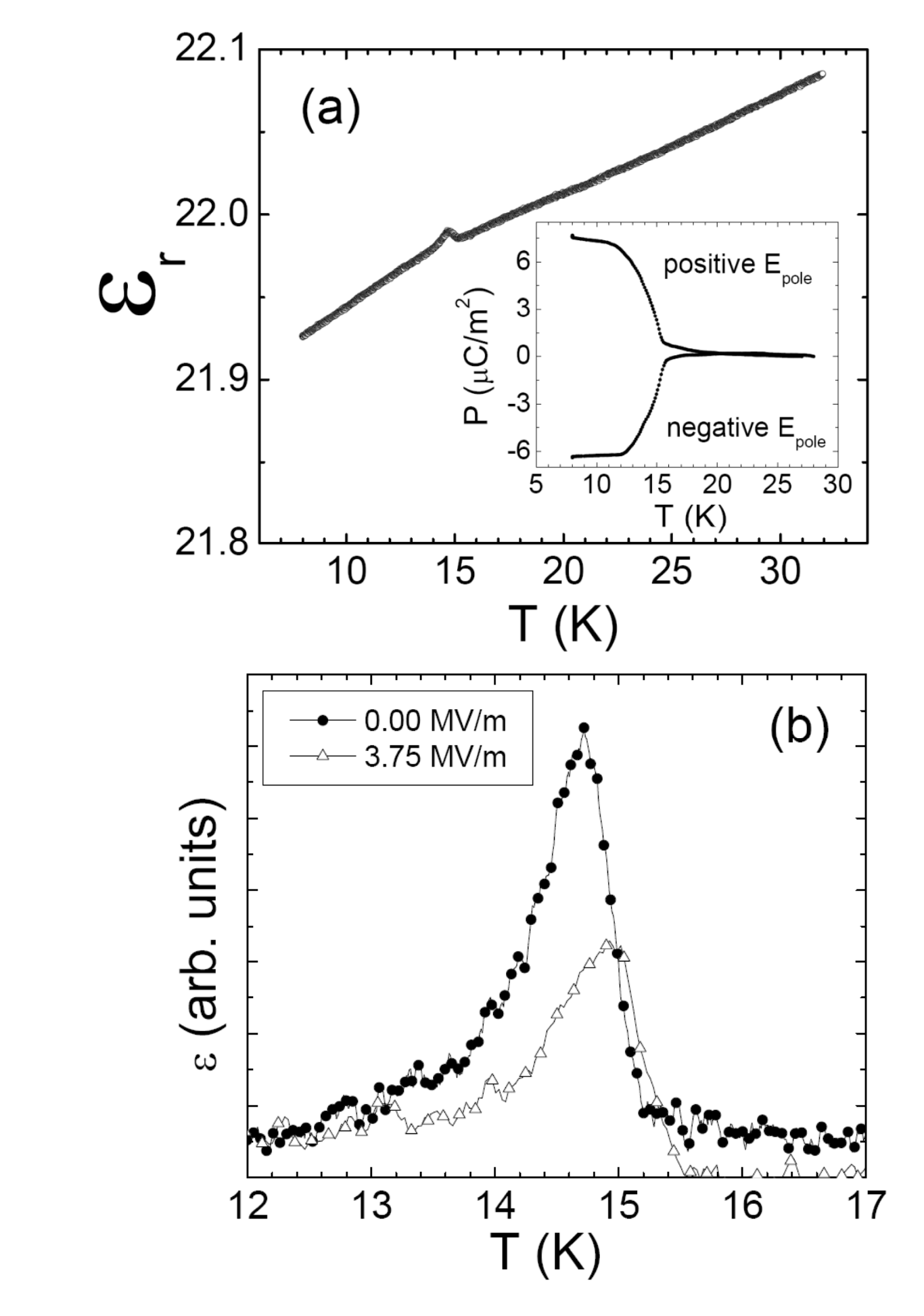}\\
\caption{(a) Temperature dependence of dielectric constant 
for \fvo\, thin films at zero field, Inset: Zero magnetic field polarization 
for \fvo\,  thin film measured at poling fields E$_{pole}$ = $\pm$10 MV m$^{-1}$ and
 (b) Temperature dependent dielectric constant measured at 
E=0 and E=3.75 MV m$^{-1}$ (background was subtracted for clarity)}
\label{fig:Fig4}
\end{figure}

The trilinear magnetoelectric coupling that produces multiferroic order, given in Eq. 6, 
also results in an electric field ($\vec E$) dependence of the magnetic structure
through the coupling term in the free energy $\Delta F = - \vec P \cdot \vec E$. We
first confirmed that these thin film samples were also multiferroic, through
measurements of the dielectric constant and pyrocurrent, illustrated in
Fig. 4(a).  The dielectric constant for the thin film \fvo\,is slightly
higher than that found for the ceramic sample.  We attribute this discrepancy
mainly to the uncertainty in accurately determining the geometrical factor
for these thin films.  The dielectric response for these thin film samples
is approximately independent of measuring frequency and the loss for these
films is tan $\delta \approx$ 0.01, which may be due to the presence of 
pinhole defects in the thin film sample as seen in the SEM micrograph in
Fig. 1(b).  The zero-field temperature dependent dielectric constant, 
measured at $f$=30 kHz, is plotted in Fig. 4(a).  There is a sharp peak
near $T_{N2}$=15 K, associated with the development of ferroelectric order
in these thin film samples.  We note that, unlike the measurements on 
bulk \fvo\,shown in Fig. 2(a), the background dielectric constant for 
\fvo\,decreases monotonically with decreasing temperature.  This behaviour
can be associated with the much larger conductivity of the thin film sample,
arising from the presence of the pinhole defects, which obscures the low
temperature increase in dielectric constant observed in bulk \fvo\,(Fig. 2(a)).  

We confirmed that the low temperature phase of the \fvo\,thin film is 
ferroelectric by integrating the pyrocurrent after poling at positive and
negative fields to yield the spontaneous polarization.  These results are
shown in the inset to Fig. 4(a) and indicate a spontaneous polarization of
6 $\mu$C/m$^2$, consistent with previous measurements on polycrystalline
bulk \fvo\,[\onlinecite{Dixit2009}]  Measurements of the dielectric response for \fvo\,thin
films under applied magnetic fields (not shown) yield a suppression of the
multiferroic transition temperature very similar to that observed in bulk 
\fvo\,(see Fig. 2(b)).

To probe the electric field control of the multiferroic phase transition
temperature, expected from the nature of the magnetoelectric coupling, 
we measured the temperature dependent dielectric response in the \fvo\,thin
film sample as a function of bias voltage.  Focusing on thin film samples
allows the application of relatively large electric fields (on the order
of MV/m) with small applied bias voltages.  We chose to probe the transition
through dielectric measurements as the magnetic anomaly at $T_{N2}$ cannot be
clearly discerned in these thin film samples.  We plot the temperature 
dependent dielectric constant measured at E=0 and E=3.75 MV/m in Fig. 4(b).
With the application of an electric field, the dielectric peak shifts 
upwards in temperature, by approximately 0.25 K in a field of E=3.75 MV/m.
We note that any sample heating, which is expected to be negligible in any
case because of the low dissipation, would raise the sample temperature
relative to the thermometer temperature, leading to an apparent decrease
in transition temperature, rather than the increase seen in Fig. 4(b).
This increase in transition temperature is consistent with an external
electric field promoting the development of ferroelectric order, and is
similar to what has been observed previously in 
multiferroic \nvo\, films [\onlinecite{Kharel2009}]. 
The relatively small increase of the ferroelectric transition under such
large applied electric fields can be directly attributed to the very small
polarization in \fvo.  We confirmed that the dielectric anomaly in Fig. 4(b)
can still be associated with the multiferroic transition, even in the 
presence of an electric field, by measuring the response under the 
simultaneous application of magnetic and electric fields (not shown). 
Although the dielectric peak broadens considerably, the continuing presence
of a single peak under such crossed fields is strong evidence that this 
anomaly reflects the multiferroic transition in \fvo.

\subsection{Magnetoelectric phase diagram for \fvo}

\begin{figure}
\centering\includegraphics[width=15cm]{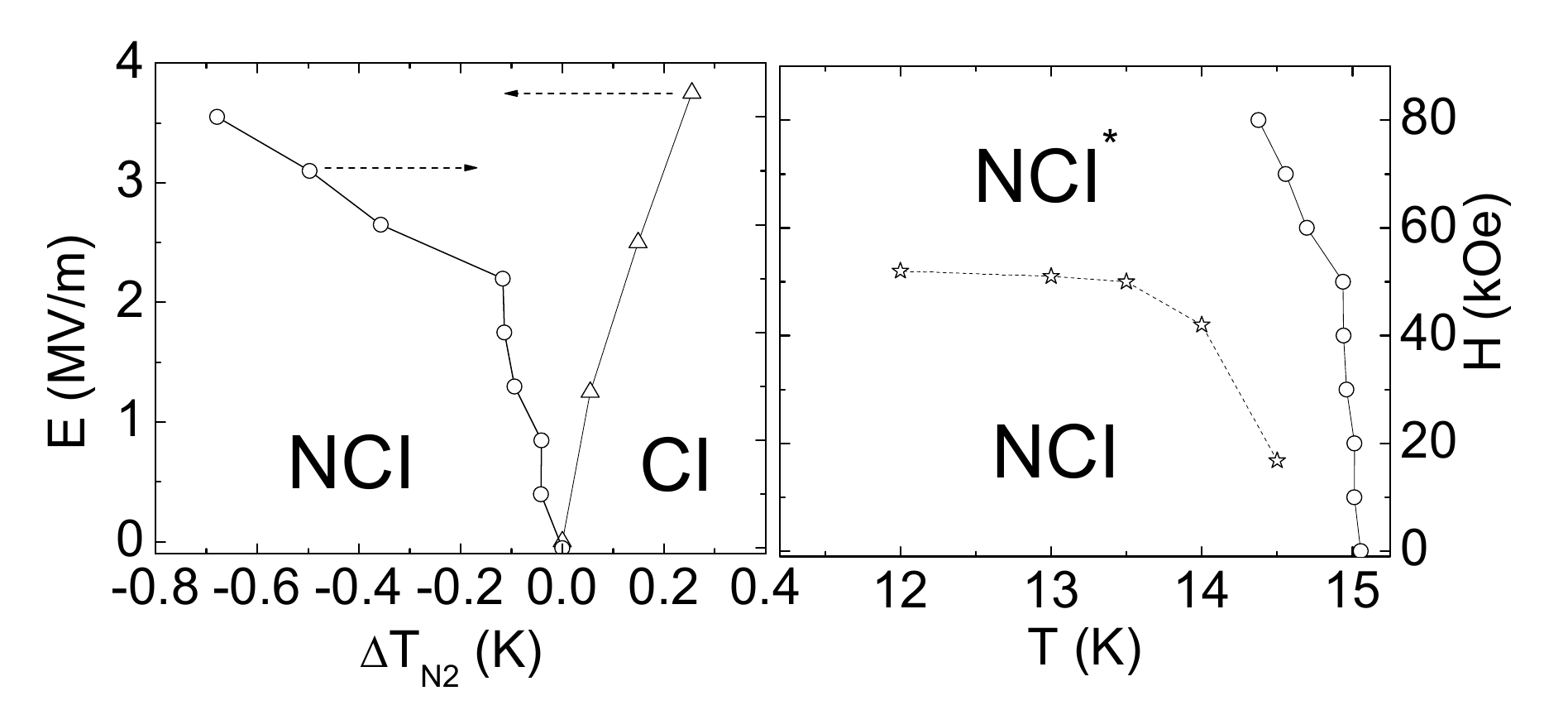}\\
\caption{(a) Electric and magnetic field dependence of the multiferroic transition
 temperature $T_{N2}$ and (b) Magnetic field dependence of 
 multiferroic transition temperature together with the proposed magnetic field 
 induced spin reorientation cross over.  Here, NCI$^*$ indicates the
 proposed phase having a spin reoriented structure}
\label{fig:Fig5}
\end{figure}

We summarize the results of these magnetoelectric and magnetodielectric 
studies on \fvo\,in Figs 5(a) and 5(b).  We plot the E-field and H-field
dependence of the multiferroic transition temperature in \fvo\,($T_{N2}$)
in Fig. 5(a), where CI and NCI represent the incommensurate magnetic 
structures below $T_{N1}$ and $T_{N2}$ respectively. This transition 
temperature is monotonically suppressed in an applied magnetic field, 
decreasing by approximately 0.7 K in an applied field of $H$=80 kOe.
  The
transition temperature, however, increases systematically with increasing
bias voltage, shifting upwards by 0.25 K in an electric field of roughly
4 MV/m. As discussed in Ref. [\onlinecite{Kharel2009}], the magnetic
field dependence of the CI-NCI phase
boundary is expected to follow $\Delta T_N \propto H^{1/2}$,
while the electric field dependence should be 
$\Delta T_N\propto E^{1/(\beta+\gamma)}$.
  This ability to control the transition temperature using either
magnetic or electric fields is a key feature for a number of proposed 
applications for multiferroic materials.  Although the size of the transition
temperature shifts in \fvo\,are likely too small to be of any practical
use, these results, taken in conjunction with previous studies on \nvo\, thin
films, provide important evidence that this behaviour is generic among 
multiferroic materials.

The magnetodielectric coupling in \fvo\,allows us to tentatively identify
the onset of short-range magnetic correlations, as indicated by the increase
in dielectric constant below $T$=30 K in Fig. 2(a), and also to propose the
onset of a spin reorientation crossover, based on the field dependent 
dielectric anomalies in Fig. 3.  Using the data from Fig. 3, we plot this
proposed spin reorientation cross-over boundary line in Fig. 5(b), together
with the magnetic field dependence of $T_{N2}$ (similar to that shown in
Fig. 5(a)).  The high-field putative spin-reorientation structure is 
labeled as NCI$^*$.  As the two boundaries do not coincide, the dielectric
anomalies in Fig. 3 are not likely to be associated with the $T_{N1}$ to
$T_{N2}$ magnetic transition, but may potentially be attributed to a change
in magnetic structure.  Magnetic field dependent specific heat measurements
(not shown) do not show any additional anomalies at this proposed cross-over,
suggesting there is a negligible change in entropy between the two 
spin structures.

\section{Conclusions}

We have presented a model for the development of multiferroic order in \fvo\,
in the context of Landau theory, which we have used to develop
constraints on the possible magnetic structures based on
symmetry considerations alone.  One of the
noteworthy predictions of this model is that the electric polarization
in the multiferroic phase is able to develop along any direction.  To further
investigate the higher order magnetoelectric coupling in this system,
we have investigated the ferroelectric and dielectric response in \fvo\,to
applied magnetic and electric fields.  The multiferroic phase transition
temperature can be tuned by applying electric or magnetic fields, in line
with the predicted trilinear magnetoelectric coupling. We find evidence for a
shift in dielectric constant well above the magnetic transition temperature
$T_{N1}$, which is expected to develop from a fourth order magnetoelectric
coupling term when short range spin correlations develop in the 
paramagnetic phase.  The dielectric constant shows a small, but distinct, increase
below the first magnetic order transition, which is consistent with the contribution from
a trilinear magnetoelectric coupling term.  We find evidence for magnetic field induced
dielectric anomalies in the non-collinear incommensurate magnetic phase of
\fvo, which we attribute to a spin reorientation transition that does not
suppress the ferroelectric structure.  These studies on \fvo\,demonstrate
the rich spin-charge coupling present in many multiferroic materials,
emphasize the importance to considering higher order expansions of
the magnetoelectric coupling to adequately explain the properties
of these materials,  and
illustrate how dielectric spectroscopy can be a valuable tool for probing
the magnetic structures in such systems.  

\section{Acknowledgments}

This work was supported by the NSF through DMR-0644823.

\begin{appendix}
\section{Landau Theory}

As discussed in detail in Ref. \onlinecite{ABHPRB} the Fourier transform of the
distribution just below a continuous magnetic ordering transition is
proportional to the critical eigenvector of the inverse susceptibility
matrix. (The critical eigenvector is the one whose eigenvalue first
approaches zero, i. e. which first becomes unstable, as the temperature
is lowered through the ordering transition.) We introduce the inverse
susceptibility as follows. The thermally averaged
spin at the site at position $\tauv$ in the unit cell at $\vec R$, 
$\langle \vec S(\vec R , \tauv) \rangle$ is defined as
\begin{eqnarray}
\langle \vec S ( \vec R , \tauv ) \rangle &
\equiv & {\rm Tr} [ \rhov \vec S_{\rm op} (\vec R , \tauv ) ] \ ,
\end{eqnarray}
where $\vec S_{\rm op}(\vec R , \tauv)$ is the quantum spin operator at
site $\vec R + \tauv$ and $\rhov$ is the density matrix:
\begin{eqnarray}
\rhov &=& \exp (- \beta {\cal H}) / [{\rm Tr} (\exp(-\beta {\cal H})]\ ,
\end{eqnarray}
where ${\cal H}$ is the Hamiltonian of the system.  The Fourier
transform of the spin distribution is given by
\begin{eqnarray}
\vec S (\vec q, \tau ) &=& N^{-1} \sum_{\vec R} \langle \vec S (\vec R , \tau)
\rangle e^{i \vec q \cdot \vec r} \  ,
\end{eqnarray}
where $\vec r$ is the actual position $\vec R + \vec \tau$ of the spin and
$N$ is the total number of unit cells in the system.
Following Landau, we write the free energy, $F$ as an expansion in powers of
$\vec S({\vec q}, \tau )$ as
\begin{eqnarray}
F &=& \frac{1}{2} \sum_{\vec q , \alpha , \beta , \tau , \tau'}
F_{\alpha , \tau ; \beta, \tau' } (\vec q) S_\alpha ({\vec q}, \tau )^*
S_\beta ({\vec q}, \tau' ) + {\cal O} [ S({\vec q})^4] \ ,
\label{FEQ} \end{eqnarray}
where the matrix ${\bf F}$ is the Hermitian inverse susceptibility matrix.
Of course, we do not know or wish to consider the exact form of ${\cal H}$
and we do not attempt to construct the inverse
susceptibility from first principles.  But we can analyze how symmetry
influences the structure of the inverse susceptibility. In what follows
we assume that the wave vector $\vec q$ at which ordering occurs
has been established experimentally and therefore we focus only on that
wave vector.

We now consider the case of FeVO$_4$ which hase six spin sites within the
unit cell of the space group P$\overline 1$.  The only point group symmetry
element is spatial inversion about the origin ${\cal I}$, so that the six
sites consist of three pairs of sites $\vec r_n$ and $\vec r_{n+3}=-\vec r_n$,
with $n=1,2,3$. Since the group of the wave vector contains only the
identity element, the standard analyses based on this group would indicate
that an allowed spin distribution function is a basis function of the
identity irrep and therefore that symmetry places no restriction on the
form of the spin distribution function.
However, since ${\cal I}$ is a symmetry of the system when all the
spins are zero, the free energy of the system for a configuration
with an arbitrary distribution of $\vec S_\alpha (\vec q , \tau)$ is
the same as that for a configuration obtained by inversion applied to
the distribution $\vec S_\alpha (\vec q, \tau)$.  So we consider
the effect of inversion on $\vec S_\alpha (\vec q, \tau)$.  The effect of
${\cal I}$ is to move a spin, without changing its orientation (because
spin is a pseudo vector), from an initial location $\vec r$, to a final
location $- \vec r$.  This  means that 
\begin{eqnarray}
{\cal I} \langle \vec S({\vec R}, \tauv) \rangle =
\langle \vec S(-{\vec R}, \overline \tauv) \rangle \ ,
\end{eqnarray}
where, for $n=1,2,3$, 
\begin{eqnarray}
\overline \tauv_n = - \tauv_n
= \tauv_{n+3} \equiv \tauv_{\overline n} \ , \hspace {0.5 in}
\overline \tauv_{n+3} = - \tauv_{n+3} = \tauv_{n}
\equiv \tau_{\overline{n+3}} \ .
\end{eqnarray}
It then follows that
\begin{eqnarray}
{\cal I} \vec S ( \vec q , \tau) &=& \vec S( \vec q , \overline \tau )^* \ .
\label{EQ5} \end{eqnarray}
 
Because we have 6 spins in the unit cell each having three Cartesian spin
components the matrix ${\bf F}$ is an 18 $\times$ 18 matrix which we write
in terms of 9 $\times$ 9 submatrices (for $n=1,2,3$ and $n=4,5,6$, 
respectively) as
\begin{eqnarray}
{\bf F} &=& \left[ \begin{array} {c c} {\bf A} & {\bf B} \\
{\bf B}^\dagger & {\bf C} \\ \end{array} \right] \ .
\end{eqnarray}
Now we consider the invariance of the free energy under spatial inversion:
\begin{eqnarray}
F &=& \frac{1}{2} \sum_{\vec q , \alpha , \beta , \tau , \tau'}
F_{\alpha , \tau ; \beta, \tau' } (\vec q) S_\alpha ({\vec q}, \tau )^*
S_\beta ({\vec q}, \tau' ) + {\cal O} [ S({\vec q})^4] \nonumber \\
&=& \frac{1}{2} \sum_{\vec q , \alpha , \tau , \beta , \tau'}
F_{\alpha , \tau ; \beta, \tau' } (\vec q) 
[ {\cal I} S_\alpha ({\vec q}, \tau )^*]
[ {\cal I} S_\beta ({\vec q}, \tau' )] + {\cal O} [ S({\vec q})^4] \nonumber \\
&=& \frac{1}{2} \sum_{\vec q , \alpha , \tau , \beta , \tau'}
F_{\alpha , \tau ; \beta, \tau' } (\vec q) 
S_\alpha ({\vec q}, \overline \tau )
S_\beta ({\vec q}, \overline \tau' )^*
+ {\cal O} [ S({\vec q})^4] \nonumber \\ &=& \frac{1}{2} 
\sum_{\vec q , \alpha , \tau , \beta , \tau'}
F_{\alpha , \tau ; \beta , \tau' } (\vec q)^*
S_\alpha ({\vec q}, \overline \tau )^*
S_\beta ({\vec q}, \overline \tau' )
+ {\cal O} [ S({\vec q})^4] \nonumber \\ &=& \frac{1}{2} 
\sum_{\vec q , \alpha , \tau , \beta , \tau'}
F_{\alpha , \overline \tau ; \beta , \overline \tau' } (\vec q)^*
S_\alpha ({\vec q}, \tau )^* S_\beta ({\vec q}, \tau' )
+ {\cal O} [ S({\vec q})^4] \ .
\label{FEQ2} \end{eqnarray}
The next-to-last equality follows because the free energy is real. The
last equality is obtained by interchanging the roles of the dummy
variables $\tau$ and $\overline \tau$ and the roles of $\tau'$ and
$\overline \tau'$.

We now compare Eq. (\ref{FEQ}) and the last line of Eq. (\ref{FEQ2}).
Since these forms have to be equal irrespective of the values of
the $S$'s, we must have that
\begin{eqnarray}
F_{\alpha , \tau ; \beta , \tau'} ( \vec q) &=&
F_{\alpha , \overline \tau ; \beta , \overline \tau'} ( \vec q)^* \ .
\end{eqnarray}
This equality relates (for $1 \leq \tau , \tau' \leq 3$)
the submatrices ${\bf A}$ and ${\bf C}$ and (for
$1 \leq \tau \leq 3$ and $4 \leq \tau' \leq 6$)
${\bf B}$ and ${\bf B}^\dagger$.  As a result we see that 
${\bf B}^\dagger = {\bf B}^*$, so that ${\bf B}$ is
symmetric and ${\bf C}= {\bf A}^*$.  Thus
\begin{eqnarray}
{\bf F} &=& \left[ \begin{array} {c c} {\bf A} & {\bf B} \\
{\bf B}^* & {\bf A}^* \\ \end{array} \right] \ .
\end{eqnarray}
Then the eigenvectors $[\Psiv , \Phiv]$, written in terms of the nine
component vectors $\Psi$ and $\Phi$  satisfy
\begin{eqnarray}
{\bf A} \Psiv + {\bf B} \Phiv &=& \lambda \Psiv \ , \hspace{0.5 in}
{\bf B}^* \Psiv + {\bf A}^* \Phiv = \lambda \Phiv \  .
\label{MATEQ} \end{eqnarray}
For instance
\begin{eqnarray}
\Psiv &=& [ S_x(\vec q , 1), \ S_y(\vec q , 1), \ S_z(\vec q , 1), \
S_x(\vec q , 2) ,\ S_y(\vec q , 2) ,\ S_z(\vec q , 2) ,\
S_x(\vec q , 3) ,\ S_y(\vec q , 3) ,\ S_z(\vec q , 3) ] \nonumber \\
\Phiv &=& [ S_x(\vec q , 4) ,\ S_y(\vec q , 4) ,\ S_z(\vec q , 4) ,\
S_x(\vec q , 5) ,\ S_y(\vec q , 5) ,\ S_z(\vec q , 5) ,\
S_x(\vec q , 6) ,\ S_y(\vec q , 6) ,\ S_z(\vec q , 6) ] \ .
\end{eqnarray}
The second equation of Eq. (\ref{MATEQ}) can be written as
\begin{eqnarray}
{\bf B} \Psiv^* + {\bf A} \Phiv^* &=& \lambda \Phiv^* \  .
\end{eqnarray}
So if $[ \Psiv , \Phiv ]$ is an eigenvector with eigenvalue $\lambda$, then
so is $\exp( i \rho) [\Phiv^*, \Psiv^*]$. In principle, these could be
two independent degenerate eigenvectors.  But if one considers the
simple case when ${\bf A} = a {\bf E}$ and $B= b{\bf E}$, where
$a$, and $b$ are scalars and ${\bf E}$ is the unit matrix, one sees that
these two solutions are, apart from a phase factor, the same. Only
for special values of the matrices are these two eigenvectors distinct
degenerate solutions.  This is an example of an accidental degeneracy
whose existence we exclude. Therefore the condition that these two
solutions only differ by a phase factor leads to the result that
\begin{eqnarray}
\Psiv = e^{i \rho} \Phiv^* \ .
\end{eqnarray}
Thus the $n$th eigenvector is 
\begin{eqnarray}
\Lambdav_n \equiv [e^{i \rho_n} \Phiv_n^* , \Phiv_n ] =
e^{ i\rho_n /2} [ e^{i \rho_n/2} \Phiv^* , e^{-i \rho_n/2} \Phiv ] \ ,
\end{eqnarray}
which we write in canonical form as
\begin{eqnarray}
\Lambda_n & \equiv & \sigma_n(\vec q) [ \Thetav_n^* , \Thetav_n] \ 
\end{eqnarray}
where $\sigma_n(\vec q)\equiv |\sigma_n(\vec q)|\exp(i \phi_n)$ is a
complex-valued amplitude and $\Thetav_n$ is normalized:
\begin{eqnarray}
1 &=& \sum_{j=1}^9 |[\Thetav_n]_j|^2 \ .
\end{eqnarray}
Since the inverse susceptibility matrix is 18 dimensional, there are
18 eigenvectors, each of this canonical form.  We identify
$\sigma_n(\vec q)$ as the order parameter which characterizes
order of the $n$th eigenvector.  As the temperature is
lowered one such solution (which we label $n=1$) becomes critical
and at a lower temperature a second solution (which we label $n=2$)
becomes critical.  As we shall see in a moment, the magnitudes of
the associated order parameters $\sigma_n({\bf q})$ and their relative
phase are fixed by the fourth order terms in the free energy which
we have so far not considered.  Using Eq. (\ref{EQ5}) we see that
\begin{eqnarray}
{\cal I} \Lambdav_n &=& {\cal I} [\sigma_n(\vec q) \Thetav_n^*, 
\sigma_n(\vec q) \Thetav_n] = [\sigma_n(\vec q)^* \Thetav_n^* ,
\sigma_n(\vec q)^* \Thetav_n ] = \sigma_n(\vec q)^* \Lambda_n \ ,
\label{INVEQ} \end{eqnarray}
which indicates that the order parameter transforms under inversion as
\begin{eqnarray}
{\cal I} \sigma_n(\vec q) &=& \sigma_n(\vec q)^* \ .
\label{IEQ} \end{eqnarray}
Also, under spatial translation, $T_{\vec R}$, we have that
\begin{eqnarray}
T_{\vec R} \sigma_n(\vec q) &=& e^{i \vec q \cdot \vec R} \sigma_n (\vec q)\ .
\label{TEQ} \end{eqnarray}
Note that Eq. (\ref{IEQ}) does {\it not} imply that the $n$th
eigenvector is invariant under inversion {\it about the origin}.
However, as we now show, it does imply that the $n$th eigenvector
is invariant about an origin which depends on the choice of phase 
of the $n$th eigenvector.  (It is obvious that a cosine wave is 
only inversion invariant about one of its nodes which need not 
occur at the origin.)  If ${\cal I}_{\vec R}$ denotes inversion 
about the lattice vector ${\vec R}$, then we have
\begin{eqnarray}
{\cal I}_{\vec R} \sigma_n({\vec q}) &=& T_{\vec R} {\cal I} T_{- \vec R} \sigma_n
(\vec q) = T_{\vec R} {\cal I} e^{-i \vec q \cdot \vec R} \sigma_n (\vec q) \nonumber \\
&=& T_{\vec R} e^{i \vec q \cdot \vec R} \sigma_n(\vec q)^* =
e^{2i \vec q \cdot \vec R\sigma_n(\vec q)} \ .
\end{eqnarray}
Let $\sigma_n(\vec q) = \sigma_n(\vec q)| e^{i\chi}$.
Then if we choose $\vec R$ so that $\vec q \cdot \vec R = \chi$, then
\begin{eqnarray}
{\cal I}_{\bf R} \sigma_n({\vec q}) &=& \sigma_n(\vec q) \ .
\end{eqnarray}
So, Eq. (\ref{IEQ}) implies inversion symmetry about a point which can be
chosen to be arbitrarily close to a lattice point for an infinite system.

Thus the contribution to the free energy from these order parameters 
$\sigma_n(\vec q)$ at wave vector $\vec q$ can be written as
\begin{eqnarray}
F &=&  \sum_n \Biggl[ a_n (T-T_n) |\sigma_n(\vec q)|^2
+  b_n |\sigma_n(\vec q)|^4 + \dots \Biggr] 
\nonumber \\ && \ + \sum_{n < m} c_{nm} |\sigma_n(\vec q) \sigma_m(\vec q)|^2
+ \sum_{n<m} \Biggl( d_{nm} [\sigma_n(\vec q) \sigma_m(\vec q)^*]^2
+ d_{nm}^* [\sigma_n(\vec q)^* \sigma_m(\vec q)]^2 \Biggr) \ ,
\label{FFEQ} \end{eqnarray}
where translational invariance indicates that for an incommensurate
wave vector the free energy is a function of $|\sigma_m|^2$, $|\sigma_n|^2$,
$\sigma_n \sigma_m^*$, and $\sigma_n^* \sigma_m$. In writing this
free energy we have assumed that the wave vectors of $\sigma_1$
and $\sigma_2$ are locked to be the same, as discussed in Ref.
\onlinecite{MKPRB}.

The generic situation in multiferroics is that as one lowers the temperature
an order parameter $\sigma_1$ first becomes nonzero and then, at a lower
temperature, a second order parameter $\sigma_2$ becomes nonzero. In
many cases, such as Ni$_3$V$_2$O$_8$[\onlinecite{Lawes2005}] or 
TbMnO$_3$[\onlinecite{MKPRL}] $\sigma_1$ and $\sigma_2$ have different 
nontrivial symmetry.  Here all the order parameters have the symmetry
expressed by Eqs. (\ref{IEQ}) and (\ref{TEQ}).  (The  phase
$\phi_2$ of the second order parameter is fixed relative to that, $\phi_1$,
of the first order parameter by the term in $d_{12}$ in Eq. (\ref{FFEQ}).)

Finally, we consider the magnetoelectric coupling, $V$, in the free energy
which is responsible for the appearance of ferroelectricity
(for which $\vec P \not= 0$, where $\vec P$ is the electric polarization).
We write
\begin{eqnarray}
F &=& F_{\rm M} + F_{\rm E} + V \ ,
\end{eqnarray}
where $F_{\rm M}$ is the purely magnetic free energy of Eq. (\ref{FFEQ}),
$F_{\rm E}$ is the dielectric potential which we approximate as
$F_{\rm E} = (1/2) \chi_E^{-1} {\bf P}^2$, where $\chi_E$ is
the dielectric susceptibility (whose crystalline anisotropy is neglected),
and to leading order in $\sigma_n$
\begin{eqnarray}
V &=& \sum_{n,m=1}^2 \sum_\gamma 
[a_{n,m,\gamma} \sigma_n(\vec q) \sigma_m(\vec q)^*
+ a_{n,m,\gamma}^* \sigma_n(\vec q)^* \sigma_m(\vec q)] P_\gamma \  ,
\end{eqnarray}
where $n$ and $m$ label order parameter modes and $\gamma$ labels the
Cartesian component of $\vec P$.  Terms linear in $\sigma_n$ are prohibited
because they can not conserve wave vector.  Terms of order $\sigma^4$ or
higher can exist.[\onlinecite{ELBIO,BET,HKAE}] The interaction $V$
has to be inversion invariant.  Since ${\cal I}\vec P = - \vec P$ and
${\cal I} |\sigma_n|^2 = |\sigma_n|^2$, we see that the terms
with $n=m$ are not inversion invariant and hence are not allowed.  Thus
\begin{eqnarray}
V &=& \sum_\gamma [a_\gamma \sigma_1(\vec q) \sigma_2(\vec q)^*
+ a_\gamma^* \sigma_1(\vec q)^* \sigma_2(\vec q)] P_\gamma \  .
\end{eqnarray}
Using ${\cal I} P_\gamma = - P_\gamma$ and Eq. (\ref{IEQ}) we see that
inversion invariance implies that
$a_\gamma = i r_\gamma$, where $r_\gamma$ is real. Then
\begin{eqnarray}
V &=& i \sum_\gamma r_\gamma [\sigma_1(\vec q) \sigma_2(\vec q)^* -
\sigma_1(\vec q)^* \sigma_2(\vec q)] P_\gamma =
2 \sum_\gamma r_\gamma |\sigma_1(\vec q) \sigma_2(\vec q)|
\sin(\phi_2 - \phi_1) P_\gamma \ .
\label{EQRES} \end{eqnarray}
Note that there is no restriction on the direction of the spontaneous
polarization, so that all components of $\vec P$ will be nonzero.
However, if the magnetic structure is a spiral, then the arguments of
Mostovoy[\onlinecite{MOST}] can be used to predict the approximate
direction of $\vec P$.  The result of Eq. (\ref{EQRES}) is quite analogous
to that for Ni$_3$V$_2$O$_8$[\onlinecite{Lawes2005}]
or for TbMnO$_3$[\onlinecite{MKPRL}], in that it requires the two modes
$\sigma_1 (\vec q)\equiv \exp(i \phi_1) |\sigma_1(\vec q)|$ and
$\sigma_2 (\vec q)\equiv \exp(i \phi_2) |\sigma_2(\vec q)|$ to be out of phase
with one another, in other words that $\phi_1 \not= \phi_2$.  

If, as stated in Ref. \onlinecite{Daoud2009}, the eigenvector is {\it not}
inversion invariant as implied by Eq. (\ref{INVEQ}), then
one would conclude that the magnetic ordering transition is not continuous.
However, the differences between the diffraction patters of the
structure of Ref. \onlinecite{Daoud2009} and that suggested here
are subtle enough[\onlinecite{PGRPC}] that our suggested structure
seems probably the
correct one.

\end{appendix}

\end{document}